**Explosive Nucleosynthesis in Core-Collapse Type II Supernovae: Insights from new C, N, Si, and Al-Mg isotopic compositions of presolar grains**

Nan Liu[1,2*], Conel M. O'D. Alexander[3], Bradley S. Meyer[4]

Larry R. Nittler[3,5], Jianhua Wang[3], Rhonda M. Stroud[5]

[1]Institute for Astrophysical Research, Boston University, Boston, MA 02215, USA; nanliu@bu.edu

[2] Physics Department and McDonnell Center for the Space Sciences, Washington University in St. Louis, St. Louis, MO 63130, USA

[3]Earth and Planets Laboratory, Carnegie Institution for Science, Washington, DC 20015, USA

[4]Department of Physics and Astronomy, Clemson University, Clemson, SC 29634, USA

[5]School of Earth and Space Exploration, Arizona State University, Tempe, AZ 85281, USA

ABSTRACT

We report C, N, Si, and Al-Mg isotope data for 39 presolar X silicon carbide (SiC) and four silicon nitride grains – a group of presolar grains that condensed in the remnants of core-collapse Type II supernovae (CCSNe) – isolated from the Murchison meteorite. Energy dispersive X-ray (EDX) data were used to determine the Mg and Al contents of the X SiC grains for comparison with the Mg/Al ratios determined by secondary ion mass spectroscopy (SIMS). Previous SIMS studies have used O-rich standards in the absence of alternatives. In this study, the correlated isotopic and elemental data of the X SiC grains enabled accurate determination of the initial $^{26}$Al/$^{27}$Al ratios for the grains. Our new grain data suggest that (*i*) the literature data for X grains are affected to varying degrees by asteroidal/terrestrial contamination, and (*ii*) the Al/Mg ratios in SiC are a factor of two (with ±6% 1σ uncertainties) lower than estimated based on the SIMS analyses that used O-rich standards. The lowered Al/Mg ratios result in proportionally higher inferred initial $^{26}$Al/$^{27}$Al ratios for presolar SiC grains. In addition, the suppression of asteroidal/terrestrial contamination in this study leads to the observation of negative trends for $^{12}$C/$^{13}$C–$^{30}$Si/$^{28}$Si and $^{26}$Al/$^{27}$Al–$^{30}$Si/$^{28}$Si among our CCSN grains. We discuss these isotope trends in the light of explosive CCSN nucleosynthesis models, based on which we provide new insights into several non-traditional CCSN nucleosynthesis processes, including explosive H burning, the existence of a C/Si zone in the outer regions of CCSNe, and neutrino-nucleus reactions in deep CCSN regions.

*Key words*: circumstellar matter – meteorites, meteors, meteoroids – nucleosynthesis, abundances–stars: supernovae

## 1. INTRODUCTION

Core collapse Type II supernovae (CCSNe) are important in cosmic evolution because they are responsible for the creation and dissemination of many of the heavy elements that are necessary for the formation of planets and life, including C, N, and O (Kobayashi et al. 2020). CCSNe also may have played a critical role in the formation and early evolution of the Solar System by providing short-lived radionuclides (e.g., Takigawa et al. 2008) and/or isotopically anomalous material responsible for large-scale isotopic variations across the protoplanetary disk (e.g., Nie et al. 2023). Quantifying the contributions of CCSNe to these cosmic events demands accurate CCSN nucleosynthesis model predictions, which, however, are hampered by uncertainties in neutrino transport, the progenitor star properties, the hydrodynamics of the explosion, and nuclear reaction rates (Limongi and Chieffi 2003). Direct observations of isotope abundances in CCSNe are ideal for testing and constraining CCSN nucleosynthesis models, but such measurements, so far, are available for only a few radioactive isotopes such as $^{56}$Ni ($t_{1/2}$ = 6 d) and $^{44}$Ti ($t_{1/2}$ = 60 a) (e.g., Diehl 2017).

Presolar grains from CCSNe that exploded before Solar System formation, such as Type X silicon carbide (SiC) grains and silicon nitride ($Si_3N_4$) grains (hereafter both are referred to as CCSN grains), provide us with a unique opportunity to directly constrain CCSN nucleosynthesis models. The CCSN origin of X and $Si_3N_4$ grains is supported by the inferred initial presence of $^{44}$Ti along with many other diagnostic isotopic signatures, including high abundances of $^{26}$Al ($t_{1/2}$ = $7.2 \times 10^5$ a) inferred from excess amounts of the decay product $^{26}$Mg (Nittler and Ciesla 2016). Micron-sized CCSN grains have been measured in the laboratory using modern mass spectrometers for isotope ratios of many elements from C to Ba with percent-level precisions (Liu et al. 2022), thus enabling testing of CCSN models in unprecedented detail. However, several factors have constrained or compromised previous isotopic studies of CCSN grains to varying degrees, as summarized below. (*i*) Among presolar SiC grains, X grains are quite low in abundance (1–2%), and it is thus time-, cost-, and labor-intensive to locate a sufficient number of samples to provide statistically meaningful data. Presolar $Si_3N_4$ grains are even rarer, being about an order of magnitude lower in abundance than presolar X SiC grains (Nittler et al. 1995). (*ii*) Presolar CCSN grains are submicron to micron in size, making their analysis challenging. For instance, Groopman et al. (2015) demonstrated that the literature X SiC grain data suffered from terrestrial and/or asteroidal Al contamination, resulting in deviations of the derived initial $^{26}Al/^{27}Al$ data from the true compositions to varying degrees. (*iii*) Presolar CCSN grain data represent the compositions of CCSN ejecta at small scale (e.g., mm or more, depending on the gas density/pressure), and the accuracy of predicted ejecta compositions at such fine spatial resolution remains unknown, especially given the complexity in the hydrodynamics of the explosions (Müller 2020).

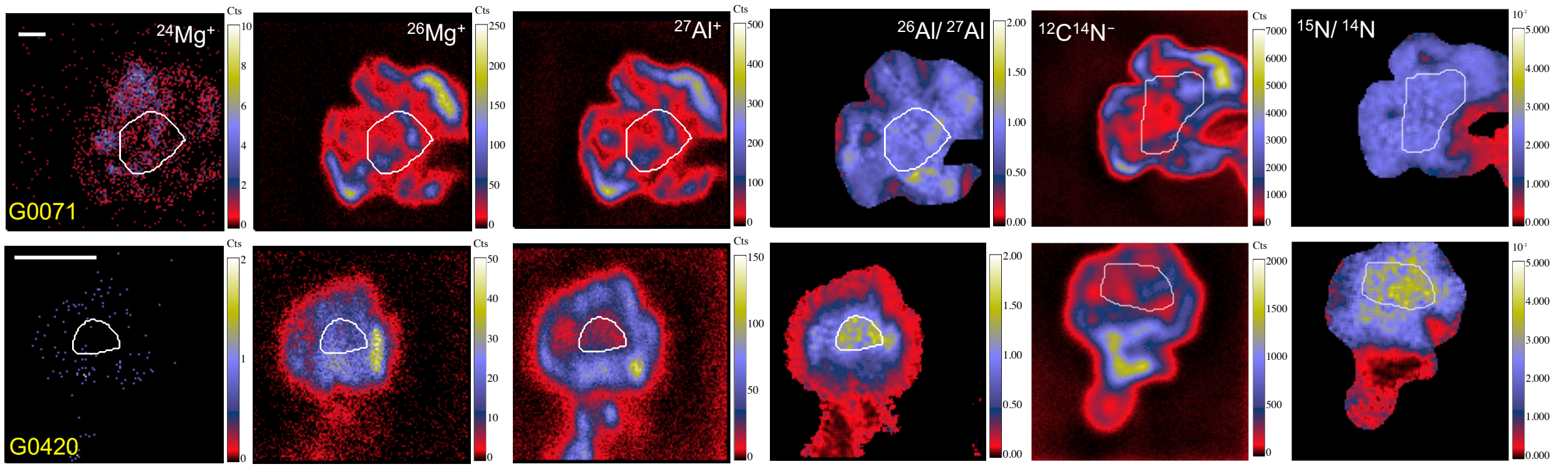

***Figure 1****. NanoSIMS ion and isotope images of two X SiC grains from this study that had relatively low (M5-A5-G0071) and high (M5-A3-G0420) levels of Al contamination. The white scalebar and contour line denotes 500 nm and the region of interest (ROI) for data reduction, respectively. The ROIs were chosen to maximally suppress N or Al contamination and/or topographic effects. The isotope ratio images were calculated using the same method as adopted for the isotope data reported in Table 1.*

In this study, we maximized the efficiency of locating CCSN grains by adopting our previously developed backscattered electron-energy dispersive X-ray (BSE-EDX) screening method (Liu et al. 2017), which led to the identification of 51 X SiC and four $Si_3N_4$ grains in total. We optimized the spatial resolution in Al-Mg isotope analyses by using a nanoscale secondary ion mass spectrometer (NanoSIMS) ion microprobe equipped with a Hyperion radio-frequency plasma $O^-$ ion source (Malherbe et al. 2016). The ~100 nm spatial resolution in all NanoSIMS ion images (Fig. 1) enabled recognition and exclusion of extrinsic contamination that had not been fully removed by presputtering[1] prior to the actual analyses. Here, we report C, N, Si, and inferred initial $^{26}Al/^{27}Al$ isotope ratios for 43 of the CCSN grains that were >500 nm. We discuss how CCSN stellar nucleosynthesis and mixing could yield the isotope trends that we observe in the data, instead of reproducing the composition of each grain by ad hoc mixing calculations as done previously. Our data-model comparison approach minimizes the effect of the bottleneck (*iii*) and can yield more accurate constraints on CCSN models.

## 2. ANALYTICAL METHODS AND CCSN MODEL SIMULATIONS

The SiC grains in this study were extracted from the Murchison (CM2) carbonaceous chondrite by means of the CsF dissolution technique described in Nittler and Alexander (2003). SiC and $Si_3N_4$ grains on Mounts #5 and #6 were first identified by automatic BSE-EDX particle analyses by following the procedure described in Liu et al. (2017). The BSE-EDX analyses were conducted with the Carnegie JEOL 6500F field emission scanning electron microscope (SEM) equipped with an Oxford Instruments silicon drift detector. The EDX data were obtained as 30 s point measurements with a 10 kV, 1 nA electron beam at the center of each particle. We focused on Mg-rich (>0.1 wt.%) SiC grains since their high Mg contents are indicative of their high initial $^{26}Al/^{27}Al$ ratios and thus their CCSN origins (Liu et al. 2017). $Si_3N_4$ grains were identified by their EDX spectra. Given the similarities in the isotopic compositions of X SiC and $Si_3N_4$ grains (Nittler et al. 1995; Lin et al. 2010), we will discuss all CCSN grain data together without distinguishing between the two phases.

We measured grains on Mount #5 and Mount #6 with the Washington University NanoSIMS 50 and Carnegie NanoSIMS 50L ion microprobes, respectively, for C, Si, and N isotopes, based on which 51 X SiC and four $Si_3N_4$ grains were identified. In all the NanoSIMS sessions, a $Cs^+$ beam of ~9 pA and ~1 pA was used for presputtering and analysis, respectively. Prior to an actual analysis, we presputtered each grain until all ion count rates became constant, which generally took 1−3 mins. During presputtering, the $^{12}C^{14}N^-$ count rate often decreased over time, suggesting the presence of surface N contamination. Any remaining contaminant that could be recognized in

[1] Presputtering refers to the process of removing a layer of material from the surface of a sample before the actual analysis begins, allowing for exposing a clean surface and reaching a state of chemical equilibrium to maintain stable count rates during the subsequent SIMS analysis.

NanoSIMS ion images were further excluded by choosing small ROIs during data reduction (Fig. 1). Among the 55 identified CCSN grains, 43 large (>500 nm) SiC and $Si_3N_4$ grains were further analyzed for their Al-Mg isotopes with the Carnegie NanoSIMS 50L ion microprobe following the procedures reported in Liu et al. (2018a). For Al-Mg isotopes, an $O^-$ beam of ~3 pA and ~1 pA was used for presputtering and analysis, respectively. Despite the often-observed Al-rich contamination rims around presolar SiC grains (e.g., G0420 in Fig. 1), we did not observe decreasing $Al^+$ count rate during presputtering, likely because the previous extensive NanoSIMS analyses had already exposed clean surfaces by removing surface Al contamination. Polished NIST 610 glass and fine-grained (1−3 μm) Burma spinel ($MgAl_2O_4$) grains were both measured as standards during the Al-Mg isotope analysis session. All the C, N, Si, and Al-Mg isotope data were collected using electron multipliers in imaging mode at a spatial resolution of ~100 nm (Fig. 1).

The initial $^{26}Al/^{27}Al$ ratio was calculated from the equation $^{26}Al/^{27}Al = [^{26}Mg_{grain} - {}^{24}Mg_{grain} \times \left(\frac{26Mg}{24Mg}\right)_{std}] / (^{27}Al_{grain} \times \Gamma_{Mg/Al})$, in which $\Gamma_{Mg/Al}$ denotes the SIMS Mg/Al relative sensitivity factor. The $\Gamma_{Mg/Al}$ value was determined to be 1.34±0.18 (1σ) and 1.22±0.08 based on NanoSIMS analyses of the NIST 610 and spinel standards, respectively. Based on NIST 610, we determined the $\Gamma_{Mg/Si}$ and $\Gamma_{Al/Si}$ values to be 5.31±0.57 and 3.96±0.31, respectively. The errors in the 610-derived values are dominated by uncertainties in the elemental abundances reported in the literature[2] (Pearce et al. 1997). The C, N, Si, and initial $^{26}Al/^{27}Al$ isotope data are reported in Table 1 with 1σ errors. Here, we will focus on these isotope systems to illustrate the necessary role of supernova core material in shaping the isotopic compositions of CCSN grains.

The EDX data were quantified with Oxford Aztec routines using a suite of pure elemental and oxide standards (see Liu et al. 2017 for details). To investigate differential X-ray absorption effects on particle EDX measurements, we performed Monte Carlo simulations of the X-ray yields for SiC with wt.% levels of Mg and Al with the CASINO 2.51 software (Fig. A1 in Appendix A).

[2] We adopted the overall average values from Table 1 of Pearce et al. (1997) for the composition of NIST 610. The calculated atomic Mg/Si, Al/Si, and Mg/Al ratios of NIST 610 glass standard are $(1.40\pm0.15)\times10^{-3}$, $(3.43\pm0.27)\times10^{-2}$, and $(4.08\pm0.53)\times10^{-2}$, respectively.

For comparison with the CCSN grain data from this study, we will adopt the set of CCSN nucleosynthesis calculations reported in Liu et al. (2018b). These calculations explored explosive nucleosynthesis based on the 25 $M_\odot$ presupernova star model of Rauscher et al. (2002) and the explosion model and reaction network of Bojazi and Meyer (2014). The calculations were run at explosion energies, *E,* ranging from $1 \times 10^{51}$ erg to $5 \times 10^{51}$ erg, and did not include neutrino-nucleus interactions. In this study, we repeated the nucleosynthesis calculations by implementing neutrino reactions (with neutrino-nucleus reaction rates from Meyer et al. 1998a) because we are interested in deep CCSN layers that are close to the collapsing stellar core, where the neutrino flux is high enough to affect the nucleosynthesis of interest. We also extended *E* to $10 \times 10^{51}$ erg in both sets of calculations. We will refer the two sets of calculations as ν- and no-ν-models. We will discuss the nucleosynthesis calculations by dividing the CCSN layers into different zones that are labeled by the two most abundant elements that they contain (Meyer et al. 1995).

## 3. RESULTS AND DISCUSSION

### 3.1 Calibrating $\Gamma_{Mg/Al}$ in Presolar SiC

An accurate determination of the initial $^{26}Al/^{27}Al$ ratio is hampered by uncertainties in $\Gamma_{Mg/Al}$ that is commonly calibrated by measuring Burma spinel (e.g., Hoppe et al. 2023). Such an O-rich standard differs significantly from SiC and $Si_3N_4$ in the sample matrix, which raises the question whether the adopted calibration procedure is appropriate. Coordinated EDX and NanoSIMS analyses of X SiC grains provide us a unique opportunity to calibrate $\Gamma_{Mg/Al}$ in SiC more accurately for the following reasons. (*i*) Although presolar SiC grains are poor in Mg because of their high condensation temperatures (Lodders and Fegley 1995), X SiC grains are an exception because of the abundant radiogenic $^{26}Mg$ from $^{26}Al$ decay. Their high $^{26}Mg$ abundances enable direct determination of $^{26}Mg/^{27}Al$ (i.e., initial $^{26}Al/^{27}Al$) by EDX analysis that is, in principle, much less affected by matrix effects than NanoSIMS analysis. (*ii*) X grains serve as the best standard for calibrating $\Gamma_{Mg/Al}$ in presolar SiC grains since X grains closely resemble the other groups of SiC grains in trace element abundances, grain size, and grain morphology.

Our CASINO simulations (Fig. A1) confirm that differential X-ray absorption effects were negligible for our measurements, thus justifying our method of deriving elemental abundances from the EDX spectra of small particles. We calculated the ($^{26}$Mg/Si)$_{EDX}$ ratios from the particle spectra by assuming that the Mg EDX peak consists of pure $^{26}$Mg (Fig. 2a). This assumption is supported by the NanoSIMS ion images, based on which we estimate that the contributions of $^{24}$Mg and $^{25}$Mg to the Mg budgets of the grains are negligible (i.e., mostly below 1%). In contrast, we observed significant amounts of Al contamination in the Mg-Al ion images of our CCSN grains. The effect of Al contamination on the SIMS data is illustrated in the lower panels of Fig. 1 for Grain M5-A3-0420: its rim has a higher Al concentration and lower $^{26}$Al/$^{27}$Al ratio than its core, corroborating the effect of asteroidal/terrestrial Al contamination at the rim. Since our CASINO simulations show that EDX measurements with a 10 kV electron beam provide Al X-ray signals predominantly from depths of 200 nm to 600 nm below the sample surface (Fig. A1), the contributions of Al X-rays from surface contamination to the total amounts sampled by EDX measurements are somewhat suppressed.

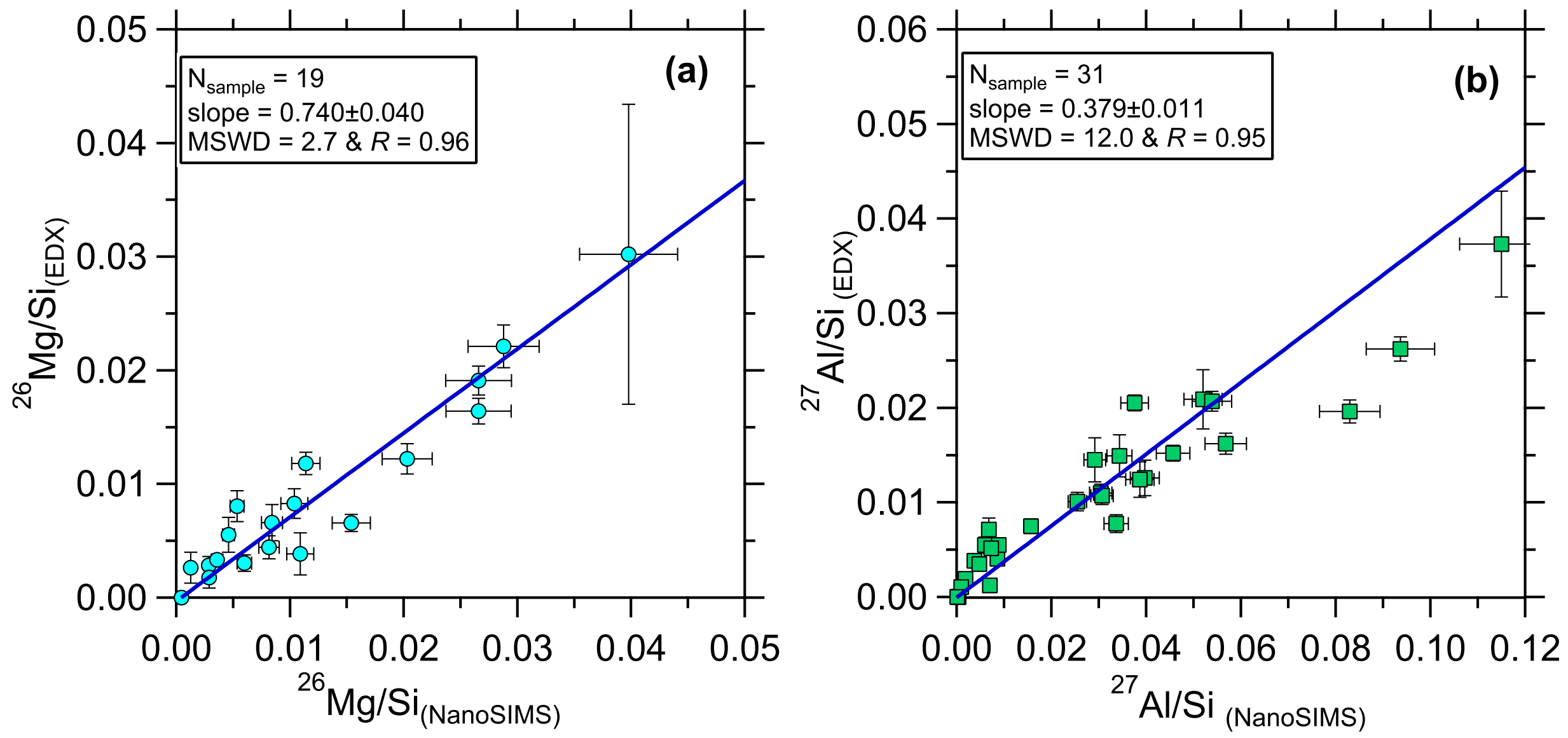

***Figure 2**. Plots comparing BSE-EDX and NanoSIMS results for X SiC (panel a) and mainstream SiC (panel b) grains. The linear fits to the grain data are plotted as solid blue lines, and the corresponding slopes are given in each panel. Errors are all 1σ. The NanoSIMS data were calculated by adopting $\Gamma_{Mg/Si}$ and $\Gamma_{Al/Si}$ values of 5.31 and 3.96, respectively. The linear fits were obtained using the CEREsFit.xlsm tool of Stephan and Trappitsch (2023). MSWD stands for mean squared weighted deviation, and R is the Pearson correlation coefficient.*

EDX and NanoSIMS analyses sample signals from a 1 $\mu m^3$ volume and a ~10 nm thin layer, respectively. A correlation is thus expected between EDX- and NanoSIMS-derived elemental ratios only if EDX and NanoSIMS analyses sampled materials from chemically homogenous grains. Therefore, we selected 19 well-isolated X SiC grains that exhibited relatively uniform $^{26}Mg^+$ distributions in their NanoSIMS ion images (denoted in Table 1). For instance, grain M5-A5-00071 was not selected because it exhibits locally enhanced $^{26}Mg^+$ by a factor of 10 within the chosen ROI (upper panels of Fig. 1), pointing to a heterogenous distribution of $^{26}Mg^+$ that is likely caused by the presence of Al-rich subgrains; in comparison, despite the significant Al contamination at its rim, grain M3-G0420 was selected because it had a relatively constant $^{26}Mg^+$ content. Figure 2a shows that given $R = 0.96$, the Mg/Si ratios of the 19 X grains based on EDX and NanoSIMS analyses are well correlated. The same set of CCSN grains, however, is poorly correlated in the case of Al/Si ratio with $R = 0.54$ and MSWD = 30.9, likely resulting from the varying degrees of Al contamination (Fig. 1) that were sampled by EDX analyses because of the poor spatial resolution (~1 μm). To suppress the effect of Al contamination, we thus analyzed a suite of large (>1 μm) mainstream (MS) presolar SiC grains, the dominant type of presolar SiC grains, on Mount #5 for their Mg-Al isotopes using NanoSIMS (Fig. A3). From the analyzed large MS grains, we selected 31 MS grains (Fig. 2b) for deriving $\Gamma_{Al/Si}$ that had the least amounts of Al contamination (i.e., no obvious Al-rich rims) and relatively homogenous Al distributions (i.e., no significant local Al enrichments). The slopes of the linear fits in Figs. 2a (0.740±0.040) and 2b (0.379±0.011) point to a true $\Gamma_{Mg/Al}$ value of 0.69±0.04,[3] which is about a factor of two lower than estimated from Burma spinel (1.22±0.08) and NIST 610 (1.34±0.18). Since the $R$ values of both linear correlations in Fig. 2 are close to unity, we infer that their large MSWD values could have been caused by the following factors. (i) We likely underestimated the uncertainties in the derived elemental ratios because, for instance, we could not accurately account for uncertainties caused by topographic effects on the NanoSIMS analyses. (ii) Given the much larger MSWD value in Fig. 2b than in Fig. 2a, other parameters such as Al contamination probably caused the larger scatter of the grain data in Fig. 2b.

[3] $\Gamma^{SiC}{}_{Mg/Al} = \Gamma^{NIST610}{}_{Mg/Al}/(0.740/0.379) = 1.34/1.95 = 0.69$.

For future studies of Al-Mg isotopes in presolar SiC grains, we recommend measuring polished NIST glass and Burma spinel standards for comparison with this study, based on which our derived $\Gamma_{Mg/Al}$ value can be applied for their initial $^{26}Al/^{27}Al$ calculations. Due to the lack of proper $Si_3N_4$ standards, we also adopted $\Gamma_{Mg/Al} = 0.69$ for deriving the initial $^{26}Al/^{27}Al$ ratios of the four $Si_3N_4$ grains. This is because the O abundance likely controls $\Gamma_{Mg/Al}$ (see Appendix A for discussion) and SiC and $Si_3N_4$ are both O-free. Figure A7 shows that the derived initial $^{26}Al/^{27}Al$ ratios of the $Si_3N_4$ grains generally follow the trend defined by X SiC grain data.

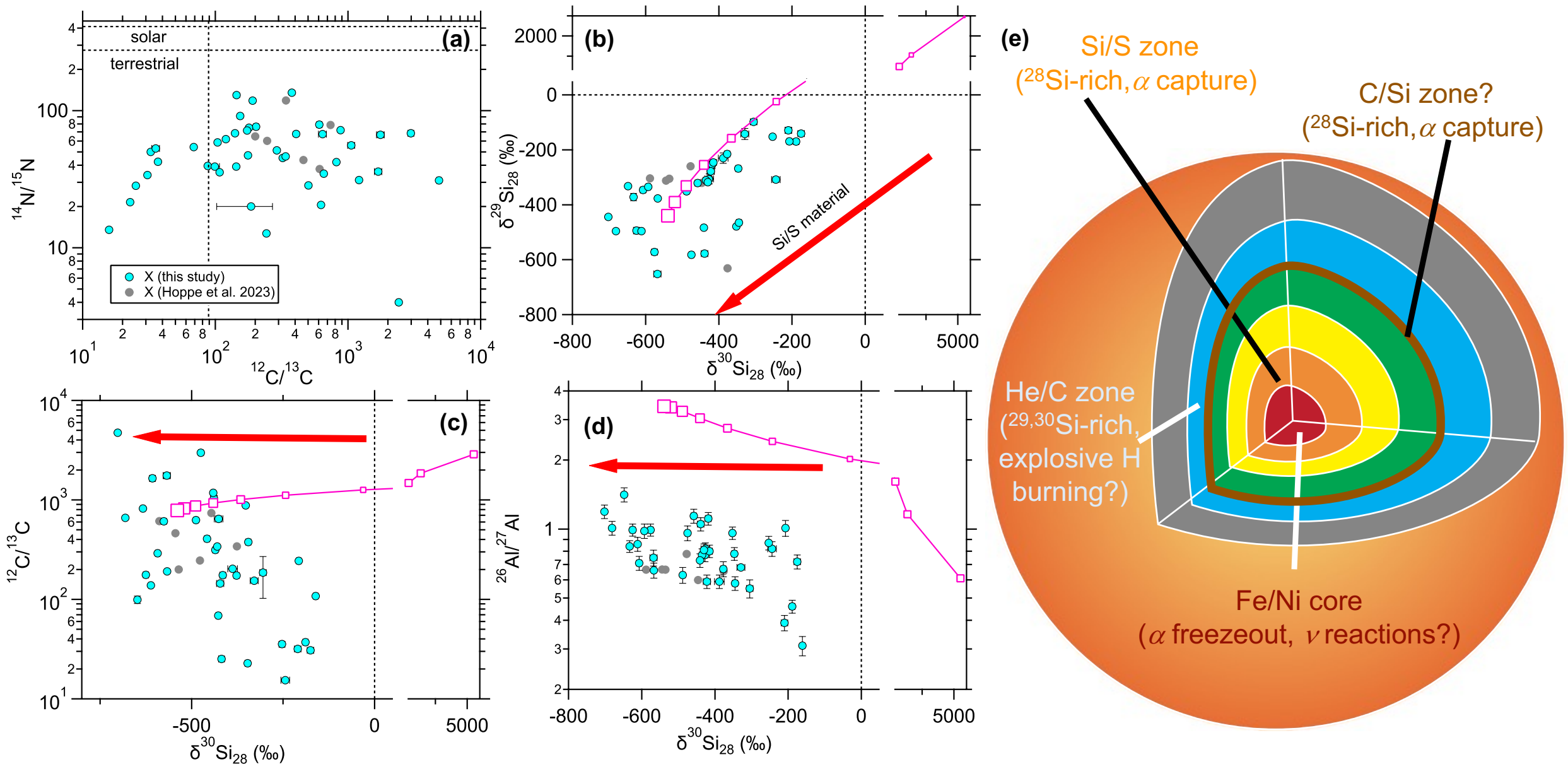

***Figure 3****. Panels (a)– (d) compare CCSN grain data from this study and Hoppe et al. (2023) with our CCSN nucleosynthesis calculations. The average composition of the Fe/Ni core was calculated by mixing Fe/Ni material from the ν- and no-ν-models with a mixing ratio of 3:2, respectively, and the calculated average compositions at explosion energies of (1–10) ×10*$^{51}$ *erg are plotted as squares (increasing size with increasing E) with lines. In panel (a), the model calculations are out of scale (see discussion in Section 3.2.4). The effects of mixing with* $^{28}$*Si-rich Si/S material are illustrated using red arrows. In panel (e), shown is a schematic diagram of the "onion-shell" internal structure of a presupernova massive star.*

### 3.2 Comparison with CCSN Nucleosynthesis Models

#### *3.2.1 Lack of C, N, and Al isotopic differences between X1 and X2 grains*

Figure 3 shows that our CCSN grains span wide ranges of C, N, and Si isotopic compositions, but their initial $^{26}Al/^{27}Al$ ratios vary only from 0.3 to 1.4, in contrast to the wide range (~0.01–0.5) observed previously (Stephan et al. 2021). Given the lowered $\Gamma_{Mg/Al}$ in SiC and observed Al contamination in this study, we suggest that the inappropriate calibration of $\Gamma_{Mg/Al}$ in previous studies caused their low upper limit[4] and that the varying degrees of Al contamination sampled in previous studies caused their wide range of initial $^{26}Al/^{27}Al$ ratios. With this new set of CCSN grain data, we observed negative trends between $^{12}C/^{13}C$, $^{26}Al/^{27}Al$, and $\delta^{30}Si_{28}$ (Fig. 3c, 3d) that are further supported by the CCSN grain data from Hoppe et al. (2023). These isotope trends are in line with the negative trends for initial $^{44}Ti/^{48}Ti$–$\delta^{29}Si_{28}$ (Lin et al. 2010) and $^{49}Ti/^{48}Ti$–$\delta^{30}Si_{28}$ values (Liu et al. 2018a).

Based on Si isotope ratios, Lin et al. (2010) proposed dividing X grains into subtypes X0, X1, and X2 grains. We, however, observed no substantial differences in $^{12}C/^{13}C$, $^{14}N/^{15}N$, and initial $^{26}Al/^{27}Al$ ratios between our X1 and X2 grains (see Fig. A7 in Appendix B). In comparison, Stephan et al. (2018) identified low $^{87}Sr/^{86}Sr$ and $^{88}Sr/^{86}Sr$ values in two X2 grains that differ significantly from those of X1 grains, and Liu et al. (2023) reported a significantly larger $^{48}Ti$ enrichment in one X2 grain than in X1 grains. Given the limited statistics, it is still questionable whether the subtype classification scheme is appropriate. We will, therefore, discuss the X grain data without the subtype classification.

*3.2.2 Underproduction of $^{26}Al$ in the C/Si zone*

That the majority of CCSN grains fall along a line in Fig. 3b points to two-endmember mixing, which suggests that CCSN grains sampled materials mainly from the innermost CCSN zones that were enriched in α isotopes (e.g., $^{28}Si$) and outer zones, i.e., the He/C zone and zones above it, that are relatively more enriched in $^{29}Si$ and $^{30}Si$ (Fig. 3e). Alternatively, Pignatari et al. (2013) ascribed the $^{28}Si$ enrichments of CCSN grains to materials from the C/Si zones (Fig. 3e) of high-explosion-energy CCSNe, where the high temperatures ($T > 3.5 \times 10^8$ *K*) enable a chain of α-capture reactions. Compared to the inner-zone scenario, the C/Si scenario requires only relatively small-scale, local mixing to reproduce CSSN grain compositions and avoids the problem of selective mixing of materials on large scales required by the former (Fig. 3e). In addition, Pignatari et al. (2015) proposed that the entire He-rich shell could retain up to percent-level H mixed inward from the surface H-rich envelope until the explosion, producing copious $^{13}$C, $^{15}$N, and $^{26}$Al by proton-capture reactions in the He/C zone during the explosion. The explosive H burning scenario greatly lowers $^{12}$C/$^{13}$C and $^{14}$N/$^{15}$N ratios and enhances $^{26}$Al/$^{27}$Al ratios in the He/C zone.

[4]Hoppe et al. (2023) adopted a $\Gamma_{Mg/Al}$ value that is half of the value determined based on their polished NIST 611 glass measurements. Their adopted $\Gamma_{Mg/Al}$ value is thus consistent with the value derived in Section 3.1, and their X grain data can be directly compared to our data as shown in Fig. 3.

The isotope trend in Fig. 3c supports the occurrence of explosive H burning in the outer He shell, given that the trend points to quite low $^{12}$C/$^{13}$C ratios (< 10–100) for the $^{30}$Si-rich endmember. It is, however, debatable whether the explosive H burning process occurred in the He/C zone as proposed by Pignatari et al. (2015) and/or in the He/N zone as shown in Liu et al. (2018b). In fact, the inferred composition of the $^{30}$Si-rich endmember is comparable to those of the He/N zone at varying *E* as reported in Liu et al. (2018b).

The isotope trend in Fig. 3d suggests that the $^{28}$Si-rich endmember tends to produce higher $^{26}$Al/$^{27}$Al ratios (> 2) than the $^{30}$Si-rich endmember. The $^{28}$Si-rich C/Si zone in the models of Pignatari et al. (2015) and Schofield et al. (2022), however, have been shown to produce $^{26}$Al/$^{27}$Al ratios that are significantly below one. Although Hoppe et al. (2023) could reproduce their X grain isotopic compositions by invoking the C/Si and explosive H burning scenarios based on the models of Pignatari et al. (2015), the explosive H burning zone ($^{30}$Si-rich) produces higher $^{26}$Al/$^{27}$Al ratios than the C/Si zone ($^{28}$Si-rich, Fig. 3e), which consequently implies a positive trend between $^{26}$Al/$^{27}$Al and $\delta^{30}Si_{28}$ that contradicts the negative trend observed in Fig. 3d. This contradiction reflects limitations and uncertainties in the ad hoc mixing calculations. Moreover, none of our nucleosynthesis calculations yielded a C/Si zone in the outer region, even though the He/C zone achieves up to $1\times10^9$ *K* at $E = 10 \times 10^{51}$ erg, thus suggesting that high temperature is not the only necessary ingredient for producing the C/Si zone. Given the strong model dependence, it is highly uncertain as to whether a C/Si zone exists in CCSNe.

*3.2.3 Evidence for the neutrino reactions in CCSNe*

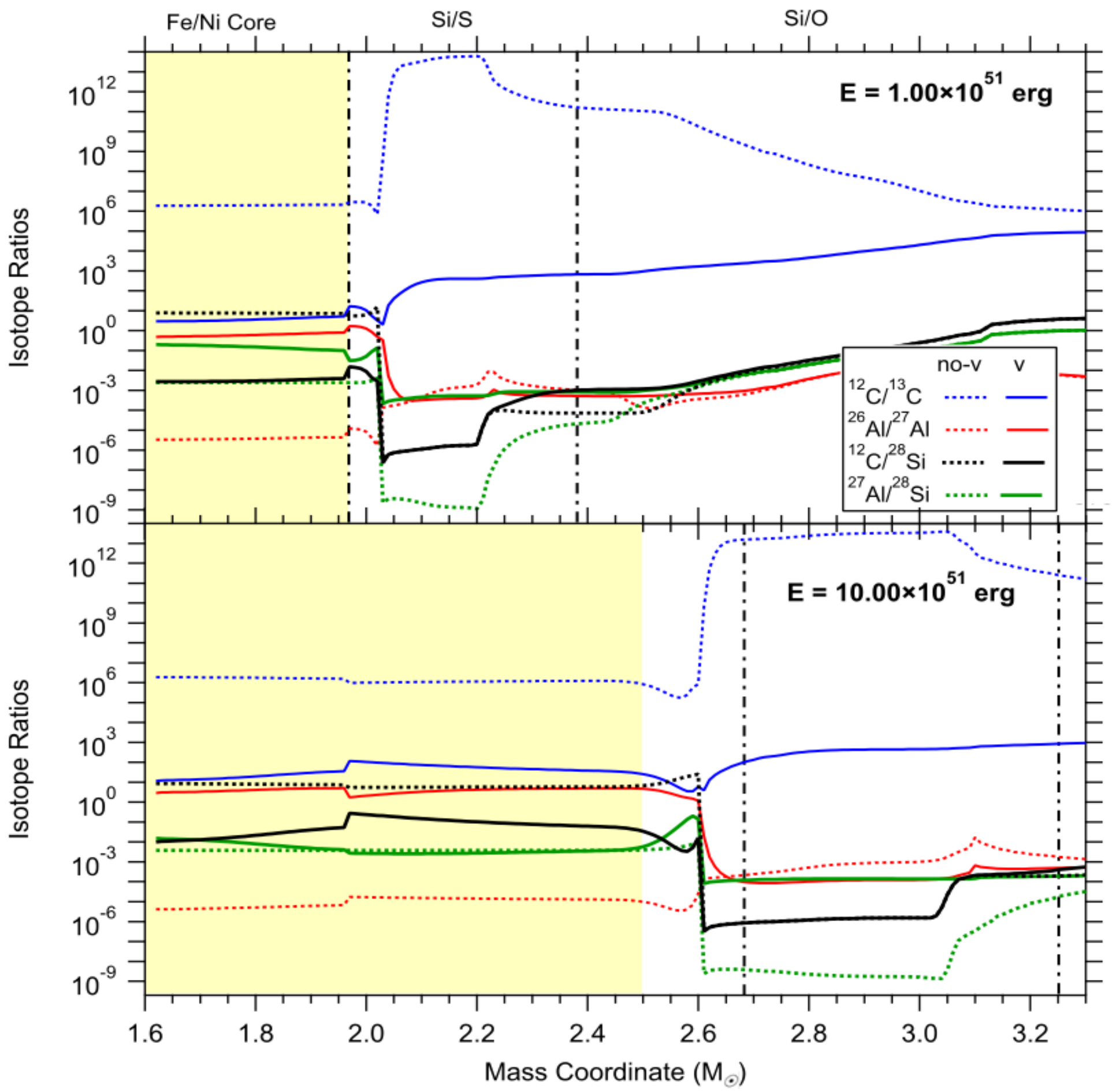

***Figure 4**. Comparison of the ν- and no-ν-models in deep CCSN regions at the lowest and highest considered explosion energies. In the Fe/Ni core, CCSN layers with relatively homogenous compositions are chosen for calculating the average compositions (highlighted in yellow) that are shown as magenta squares in Figs. 3 and A8.*

Figure 4 illustrates that neutrino reactions significantly enhance the production ratio of $^{26}Al/^{27}Al$ in the Fe/Ni core but not in the Si/S zone, which is found in our CCSN models at all explosion energies. This observation points out that the enhanced production of $^{26}Al/^{27}Al$ requires both the presence of neutrinos and an α-rich freezeout from quasi-equilibrium (see Meyer et al. 1998b for details). Neutrino-nucleus reactions during the α-rich freezeout maintain a relatively high abundance of free protons. Proton-capture reactions then enhance the production of $^{26}Al$ over the freezeout without neutrino-nucleus interactions. A detailed comparison of the CCSN grain data from this study with the $\nu$-model calculations for the Fe/Ni core is shown in Fig. A8. Figure A8 illustrates that the CCSN grain data link high initial $^{26}Al/^{27}Al$ ratios (>2) to the $^{28}Si$-rich endmember, in line with the high $^{26}Al/^{27}Al$ ratios predicted by the $\nu$-models for the Fe/Ni core at $E > 2 \times 10^{51}$ erg. Thus, the Fe/Ni core likely contributed to the CCSN ejecta from which the CCSN grains condensed.

We further illustrated the effect of mixing additional material from the adjacent Si/S zone in Fig. A8. This is because the Fe/Ni core contains pure $^{48}Ti$ and cannot account for the large $^{49}Ti$ excesses observed in X grains, which require contributions from the Si/S zone (Liu et al. 2018a). Given the copious production of pure $^{28}Si$ in the Si/S zone, this additional mixing results in simultaneous reduction in $^{29}Si/^{28}Si$ and $^{30}Si/^{28}Si$ with negligible effects on the other isotope ratios (for reproducing the isotopic compositions of CCSN grains). Thus, the composition of the mixed Fe/Ni and Si/S ejecta would approach $\delta^{30}Si_{28} = -1000‰$ along a slope-1 line in the Si 3-isotope plot by mixing with Si/S zonal material. This mixed ejecta that consists of materials from both the Fe/Ni core and Si/S zone provide a natural explanation as to why CCSN grains data do not fall along a slope-1 line (Fig. A8b) if the Si/S zone with pure $^{28}Si$ is one of the two endmembers.

This large-scale mixing between innermost CCSN material with outer C-rich material is supported by the observations that $^{56}Ni$ and $^{44}Ti$ knots in CCSN remnants are inside out, i.e., lie outside the central regions of CCSN remnants (e.g., Hwang & Laming 2012). The observed $^{56}Ni$ and $^{44}Ti$ were inferred to have been assembled under $\alpha$-rich freezeout, corresponding to the Fe/Ni core material (Wongwathanarat et al. 2017). Furthermore, based on three-dimensional model simulations, the restricted mixing of Fe/Ni and Si/S materials with those from the outer He/C zone (and zones above it), is suggested to be a natural consequence of shock deceleration in the He shell and consequent pile up of metal-rich knots from the interior in the He shell (Hammer et al. 2010; Wongwathanarat et al. 2017). The simulations further predict that Rayleigh Taylor instabilities could develop at the H/He interface and eventually lead to metals and He being mixed into the H envelope, in line with the suggestions from CCSN grain data (e.g., Xu et al. 2015).

*3.2.4 The problem of $^{12}C/^{13}C$*

In contrast to the high $^{12}C/^{13}C$ value of > 1000−10,000 inferred for the $^{28}Si$-rich endmember (Fig. A8c), $^{12}C/^{13}C$ is predicted to lie below 100 in the Fe/Ni core in the presence of neutrino reactions as shown in Fig. A8c. In a proton-rich condition, $^{13}C$ and $^{26}Al$ are generally predicted to be abundantly produced together at different stellar sites (e.g., nova explosions; José et al .2004). Therefore, the high $^{12}C/^{13}C$ and $^{26}Al/^{27}Al$ ratios inferred for the $^{28}Si$-rich endmember seem counterintuitive. Core-collapse explosions, however, are complex phenomenon that are in great contrast to other stellar environments as, for instance, $\alpha$-rich freezeouts occur solely in CCSNe. Specifically, $^{12}C$ is a product of $\alpha$-rich freezeouts, and $^{13}C$ is made abundantly in the proton-rich condition created by neutrino-nucleus reactions. Thus, the data-model discrepancy for $^{12}C/^{13}C$ could imply, for instance, uncertainties in the calculation of $\alpha$-rich freezeouts. Beside uncertainties in nucleosynthesis calculations, below we explore one scenario to enhance the high $^{12}C/^{13}C$ ratio of the core that involves mixing core materials experiencing varying proton fluxes.

It is conceivable that in the Fe/Ni core materials are contained in different bubbles that could encounter different neutrino fluxes, based on which one would expect mixing to occur between materials that experienced strong and weak proton fluxes. Although the model predictions for the $^{12}C/^{13}C$ ratio in the core is not quite sensitive to the neutrino flux across a wide range based on our model tests, the Fe/Ni core is predicted to contain almost pure $^{12}C$ in the absence of neutrino reactions (Fig. 4). In addition, three-dimensional hydrodynamic model simulations predict that the Fe/Ni core material penetrates into the He shell at different rates (e.g., Hwang & Laming 2012), which means that the Fe/Ni material probably left the deep core region at different times so that some of the material may not have time to experience the neutrino-reaction-induced proton flux, corresponding to the no-$\nu$ scenario. Also, three-dimensional hydrodynamic models predict the most fast-moving core material to have the highest likelihood to penetrate into the He shell (e.g., Hwang & Laming 2012), the condition inferred for producing the SiC and $Si_3N_4$ grains from our study.

As a test, we mixed Fe/Ni core material from the $\nu$- and no-$\nu$-model with a mixing ratio of 3:2, respectively. Our test revealed that such mixing could lead to significant increases in $^{12}C/^{13}C$ with much smaller effects on the other isotope ratios in Fig. 3, thus accounting for the large $^{12}C$ ($^{12}C/^{13}C$ > 1000–10,000) and $^{26}Al$ ($^{26}Al/^{27}Al$ > 2) enrichments in the $^{28}Si$-rich endmember. The predicted compositions at varying *E* based on this mixing scenario are shown in Fig. 3 for comparison with the grain data. The predicted $^{14}N/^{15}N$ for the Fe/Ni core increases from 0.06 to 1.00 with increasing *E*, all of which, however, are very low with respect to the CCSN grain data and are thus not shown in Fig. 3a. Given the lack of any trend observed between $^{14}N/^{15}N$ and $^{30}Si/^{28}Si$, we cannot derive any constraints on the $^{14}N/^{15}N$ ratio of the $^{28}Si$-rich endmember, which thus precludes data-model comparisons in a similar way as for the other isotope systematics. In addition, the $^{14}N/^{15}N$ ratio of the Fe/Ni core is affected by the radioactive decay of $^{14}C$ ($t_{1/2} = 5730$ a), and the relative ratio of $^{14}C/^{14}N$ that was initially incorporated into CCSN grains remains poorly constrained. Nevertheless, a positive trend between $^{12}C/^{13}C$ and $^{14}N/^{15}N$ ratios seems to exist among CCSN grains with subsolar carbon isotope ratios, likely pointing to coproduction of $^{13}C$ and $^{15}N$ in the outer CCSN regions by explosive H burning.

## 4. CONCLUSIONS

The correlated isotopic and elemental data of X grains from this study enabled accurate determination of their initial $^{26}Al/^{27}Al$ ratios. Our new grain data suggest that the Al/Mg ratios in SiC are a factor of two lower than estimated based on the SIMS analyses that used O-rich standards, corresponding to a factor of two increase in the derived initial $^{26}Al/^{27}Al$ ratios for presolar SiC grains. For future studies of Al-Mg isotopes in presolar SiC grains, we recommend measuring polished NIST glass and Burma spinel standards for comparison with this study, based on which our derived $\Gamma_{Mg/Al}$ value can be applied for their initial $^{26}Al/^{27}Al$ calculations.

For the first time, we observed negative trends for $^{12}C/^{13}C$–$\delta^{30}Si_{28}$ and $^{26}Al/^{27}Al$–$\delta^{30}Si_{28}$ among CCSN grains, which are in line with the negative trends for $^{44}Ti/^{48}Ti$–$\delta^{30}Si_{28}$ and $^{49}Ti/^{48}Ti$–$\delta^{30}Si_{28}$ reported in the literature. These isotope trends corroborate the two-endmember ($^{28}Si$-rich and $^{30}Si$-rich) mixing scenario suggested by the Si isotopic compositions of CCSN grains. The isotope trends from this study suggest lower $^{13}C$ and higher $^{26}Al$ enrichments in the $^{28}Si$-rich endmember, and higher $^{13}C$ and lower $^{26}Al$ enrichments in the $^{30}Si$-rich endmember.

The CCSN grain data do not support previous suggestions that the C/Si zone was the source of the $^{28}Si$-rich endmember, given that current CCSN stellar nucleosynthesis models all underproduce $^{26}Al$ in this zone. Instead, the grain data favor mixtures of the innermost CCSN Fe/Ni and Si/S zones as the $^{28}Si$-rich endmember. Our model calculations further illustrate that, during $\alpha$-rich freezeout in the presence of neutrinos, a proton-rich condition could be created in the core, leading to abundant $^{26}Al$ production and thus accounting for the high $^{26}Al/^{27}Al$ ratios inferred for the $^{28}Si$-rich endmember. However, the models predict $^{12}C/^{13}C$ ratios lower than inferred from the grain data for the Fe/Ni core. The data-model discrepancy could be reconciled if the core material that penetrated into the He shell experienced a wide range of neutrino fluxes.

The $^{30}Si$-rich endmember is most likely a mixture of He- and H-shell material. The grain data also suggest that explosive H burning must have occurred in the outer He-shell to produce the low $^{12}C/^{13}C$ and $^{14}N/^{15}N$ ratios of the $^{30}Si$-rich endmember. The exact location for the explosive H burning, however, is unclear, and both the He/C and He/N zones of the He-shell remain potential candidates.

The large-scale mixing of innermost CCSN material with outer He- and H-shell materials, is in line with astronomical observations and three-dimensional model simulations. However, our conclusions, which are based on C, Al-Mg, and Si isotope ratios and one-dimensional model calculations, need to be tested by including analyses of more isotope systematics in the grains and considering state-of-the-art multidimensional CCSN simulations (e.g., Sandoval et al. 2021) and their nucleosynthesis yields (e.g., Sieverding et al. 2023).

Acknowledgements: This work was supported by NASA through grants 80NSSC20K0387 to N.L., NNX10AI63G and NNX17AE28G to L.R.N, 80NSSC20K0338 to B.S.M, and 80NSSC23K0011 to R.M.S. CASINO simulations used the CASINO V2.51 software made available by the University of Sherbrooke, Montreal Canada.

## Supplementary Material

## Appendix A

## Mg/Si and Al/Si Ratios of Presolar SiC Grains

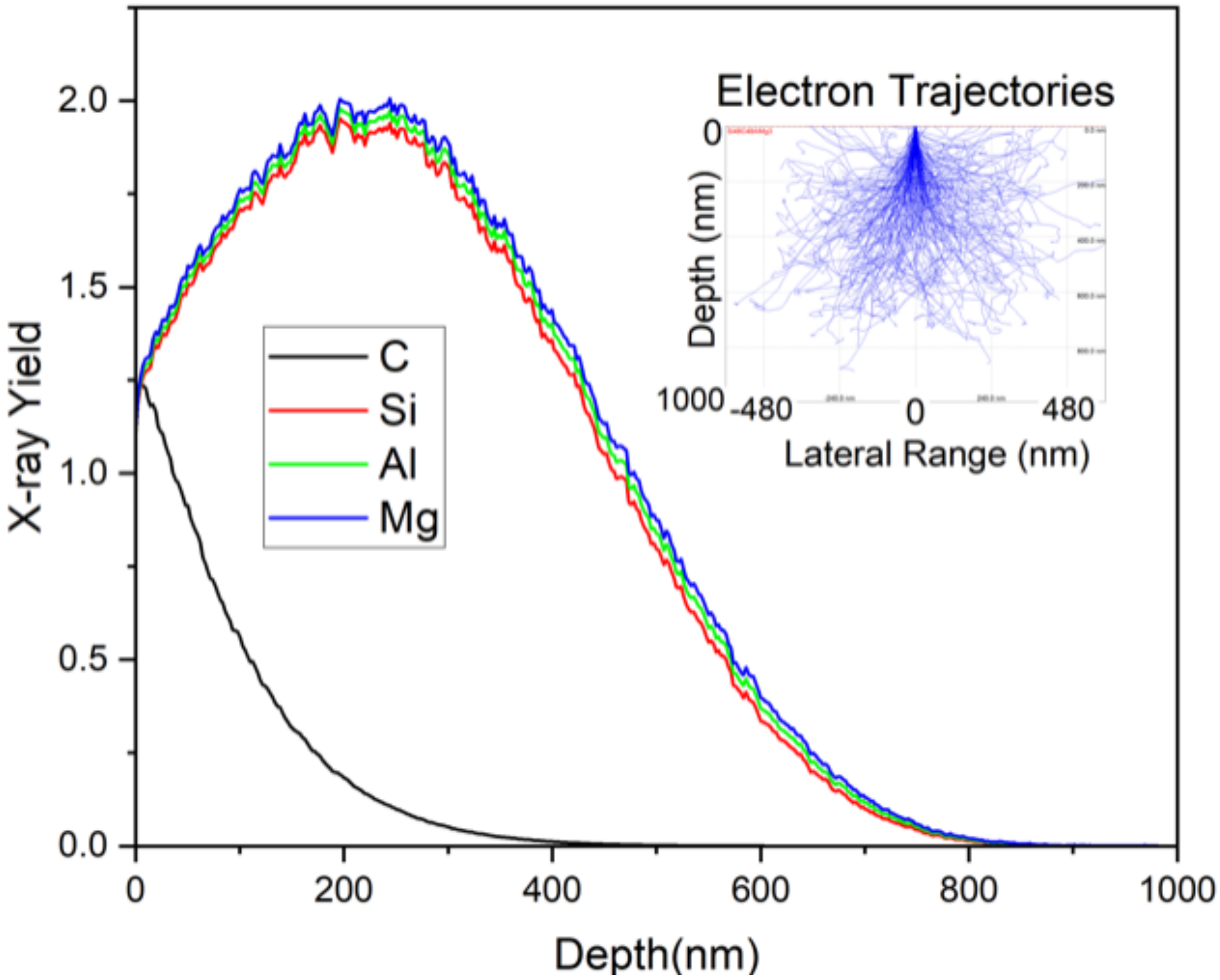

**Figure A1**. *CASINO simulation (Drouin et al. 2007) of characteristic X-ray yields (C, Si, Al, and Mg K peaks) and electron trajectories for a 1 nA, 10 kV beam of 5 nm in size incident on an infinite slab of $Si_{48}C_{48}AlMg_3$. In the inset, blue trajectories represent 1,000 of the 10,000 calculated electron trajectories within the slab. The electron trajectory distribution spans a ~0.8 μm diameter teardrop-shaped interaction volume, comparable in size to the average grain diameter for this study.*

We performed CASINO simulations for 10,000 electrons in a 5 nm, 1nA, 10 kV beam incident on an infinite slab of $Si_{48}C_{48}AlMg_3$, with a density of 3.31 g/cm$^3$. The CASINO simulation in Fig. A1 reveals nearly identical Mg, Al, and Si X-ray yield depth distribution profiles. This indicates that the X-ray yield is not appreciably affected by differential absorption, and elemental quantification using PhiRhoZ methods are valid despite the particle morphology, i.e., Mg/Si and Al/Si values are insensitive to variations in detector position or SiC particle morphology. Furthermore, the electron interaction volume (Fig. A1 inset) illustrated by 1000 of the 10,000 simulated trajectories (blue lines) is ~1 μm$^2$ lateral range and 0.8 μm in depth, which indicates that the X-ray signal samples a representative, subsurface volume of the particle for most particle sizes in this study (0.5 μm to 1 μm diameter).

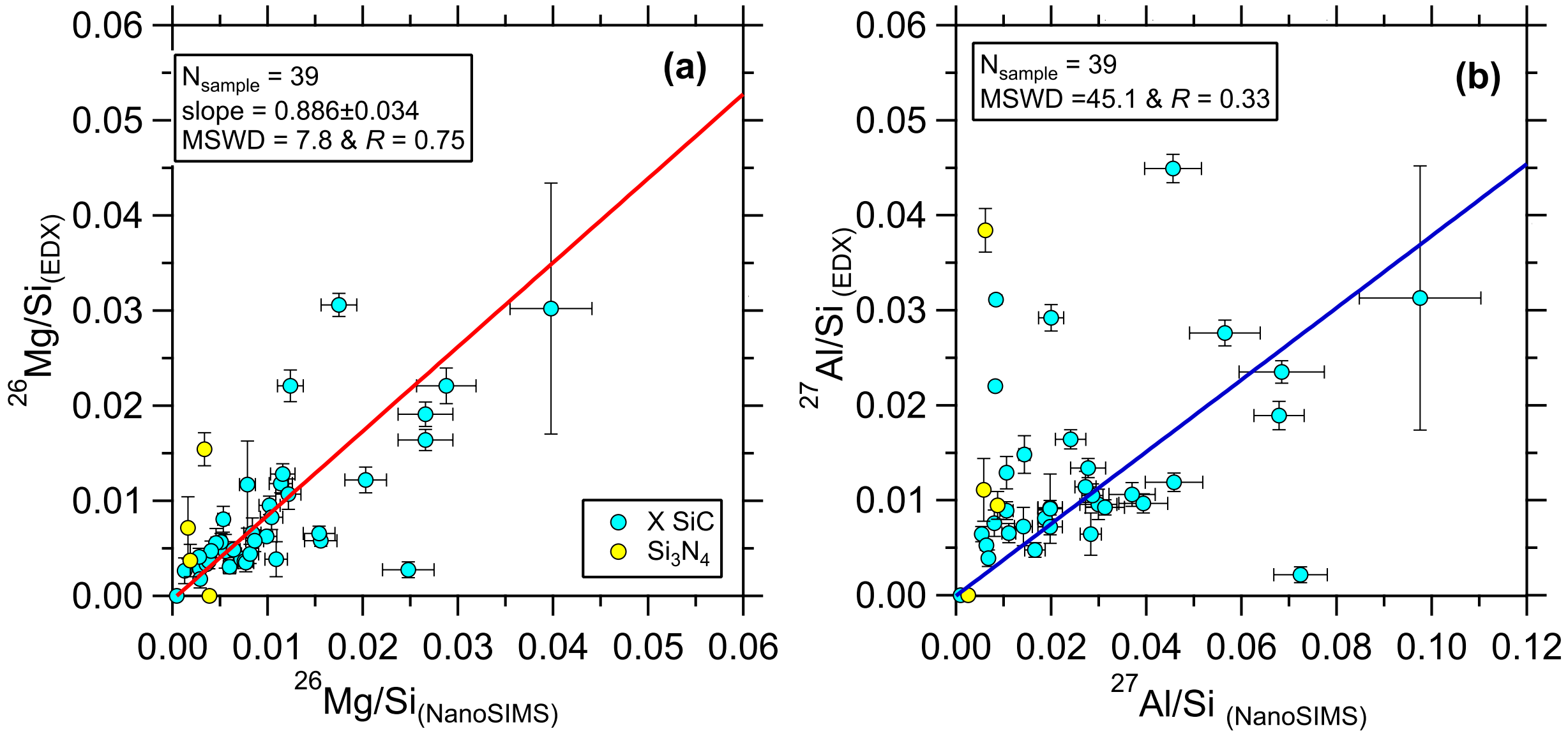

**Figure A2**. *Plots comparing BSE-EDX and NanoSIMS analysis results of all X grains from this study. In panel (a), the linear fit to the grain data is plotted as a solid red line and was obtained by using the CEREsFit.xlsm tool of Stephan and Trappitsch (2023). In panel (b), given the small R value, we did not obtain any linear fit to the grain data and instead plotted the linear fit to the selected MS grain data in Fig. 2b for comparison. Data for $Si_3N_4$ grains are plotted for comparison. Errors are all 1σ.*

Figure A2a shows that the correlation between EDX- and NanoSIMS-derived $^{26}$Mg/Si ratios determined based on the full set of our X grain data (0.89±0.03, 1σ error) agrees with that determined based on the 19 selected X grains (relatively chemically homogenous) in Fig. 2a (0.74±0.04) within 2σ uncertainties. The larger MSWD along with the smaller *R* values in the former are therefore likely caused by the heterogenous distributions of Al and thus $^{26}$Mg in 20 of the 39 X grains, which must have led to sampling materials with varying $^{26}$Mg/Si ratios from each of these grains by coordinated EDX and NanoSIMS analyses. The 39 X grains are also much less well correlated for Al/Si in Fig. A2b than for Mg/Si in Fig. A2a, given the significantly enlarged MSWD and reduced R values of the former. The differences likely result predominantly from the effects of Al contamination on EDX-derived Al/Si ratios.

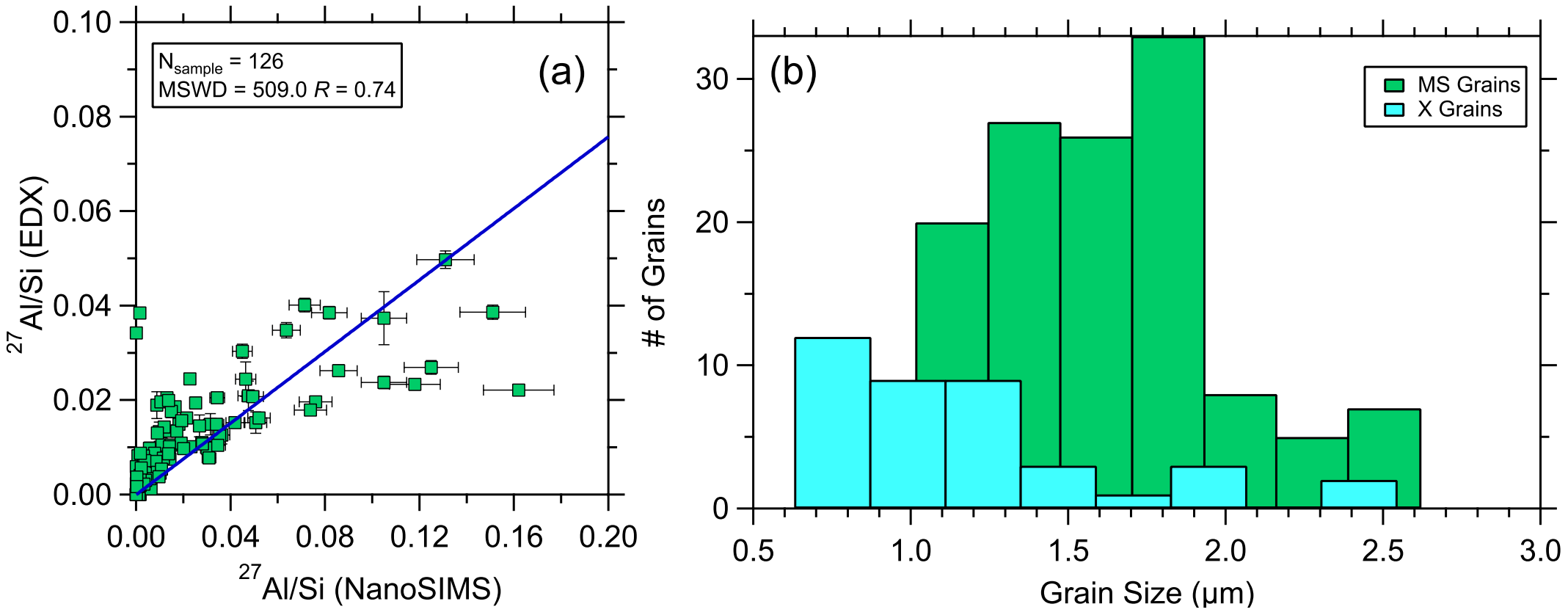

**Figure A3**. *The plot in panel (a) is the same as Fig. A2b but for 126 MS grains. In panel (b), MS grains from panel (a) are compared to X grains in Fig. A2 for their grain size distributions.*

Figure A3a further illustrates that 126 MS grains from this study are better correlated than the 39 X grains in Fig. A2b between their EDX- and NanoSIMS-derived Al/Si ratios. The better correlated grain data in the former likely result from the suppressed effects of Al contamination on their EDX data because of their enlarged sizes (Fig. A3b) and, in turn, increased volume-to-rim ratios (since Al contaminations mainly appear as rims, e.g., Grain 0420 in Fig. 1). Compared to Fig. 2b, the enlarged MSWD and reduced *R* values in Fig. A3a are likely caused by (i) sampling materials with varying Al/Si ratios by EDX and NanoSIMS analyses, given that the grains are scattered both below and above the linear fit and/or (ii) sampling higher levels of Al contaminations by EDX analyses, given that a significant fraction of the MS grains lie above the linear fit.

Figure A4 compares the initial $^{26}$Al/$^{27}$Al ratios derived based on EDX and NanoSIMS analyses. A significant fraction of the X grains from this study and Hoppe et al. (2023) are scattered around 1:1 line within 1σ uncertainties, thus corroborating our derived $\Gamma_{Mg/Al}$ (0.69±0.04) value in SiC. That several X grains lie significantly below the 1:1 line, is most likely caused by (i) the significant Al contaminations sampled by their EDX analyses (e.g., Grain M5-A3-0420) and/or (ii) dilution from adjacent grains in the EDX data due to grain aggregation (Grain M5-A7-1003).

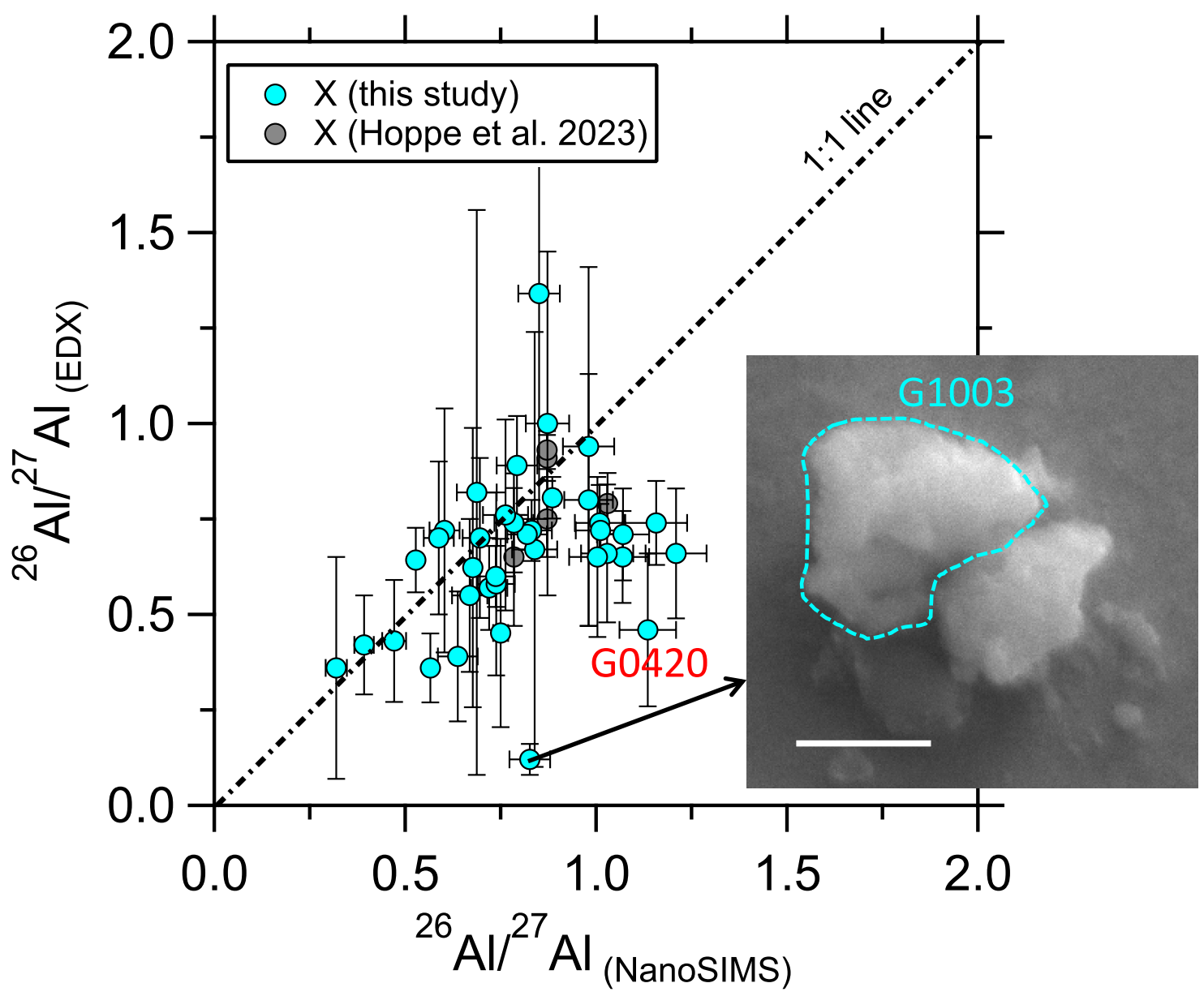

**Figure A4**. *EDX- and NanoSIMS-derived initial $^{26}Al/^{27}Al$ ratios of X grains from this study and Hoppe et al. (2023). Labeled is Grain M5-A3-0420 that had significant Al contamination (Fig. 1). In the insert, the SEM image of Grain M5-A7-1003 shows that this X grain was part of a SiC aggregate so that the adjacent SiC, which is an MS grain with $^{26}Al/^{27}Al$ = 0.001, significantly lowered its EDX-derived Mg content and, in turn, $^{26}Al/^{27}Al$ ratio. The white scalebar denotes 500 nm.*

### New $\Gamma_{Mg/Al}$: Matrix Effect or Grain Size Effect?

What is the cause of the factor of two decrease in $\Gamma_{Mg/Al}$ in SiC compared to O-rich standards such as NIST glass and Burma spinel? Two plausible explanations are: (i) SiC is O-free and thus chemically different from O-rich standards, i.e., different matrix chemistries, and $\Gamma_{Mg/Al}$ varies with matrix, and (ii) presolar SiC rains are smaller than polished NIST glass and Burma spinel, and $\Gamma_{Mg/Al}$ varies with grain size (Hoppe et al. 2023).

It is well-known that the local O abundance increases the positive secondary ion yield of many elements in SIMS, such that "oxygen flooding" wherein a low pressure of $O_2$ is leaked into the analysis chamber is a common method for increasing SIMS yields (e.g., Zalm & Vriezema 1992). Since the electronegativities of Mg, Al, and Si all differ (e.g., Fan et al. 1992), one would expect that their relative yields should depend on the O abundance. Since $O_2$ flooding is not possible in the NanoSIMS, significant differences in yields for O-free phases like SiC compared to oxides or silicate glasses would be expected, strongly supporting possibility (i). Nevertheless, it is worth examining possibility (ii) in some detail.

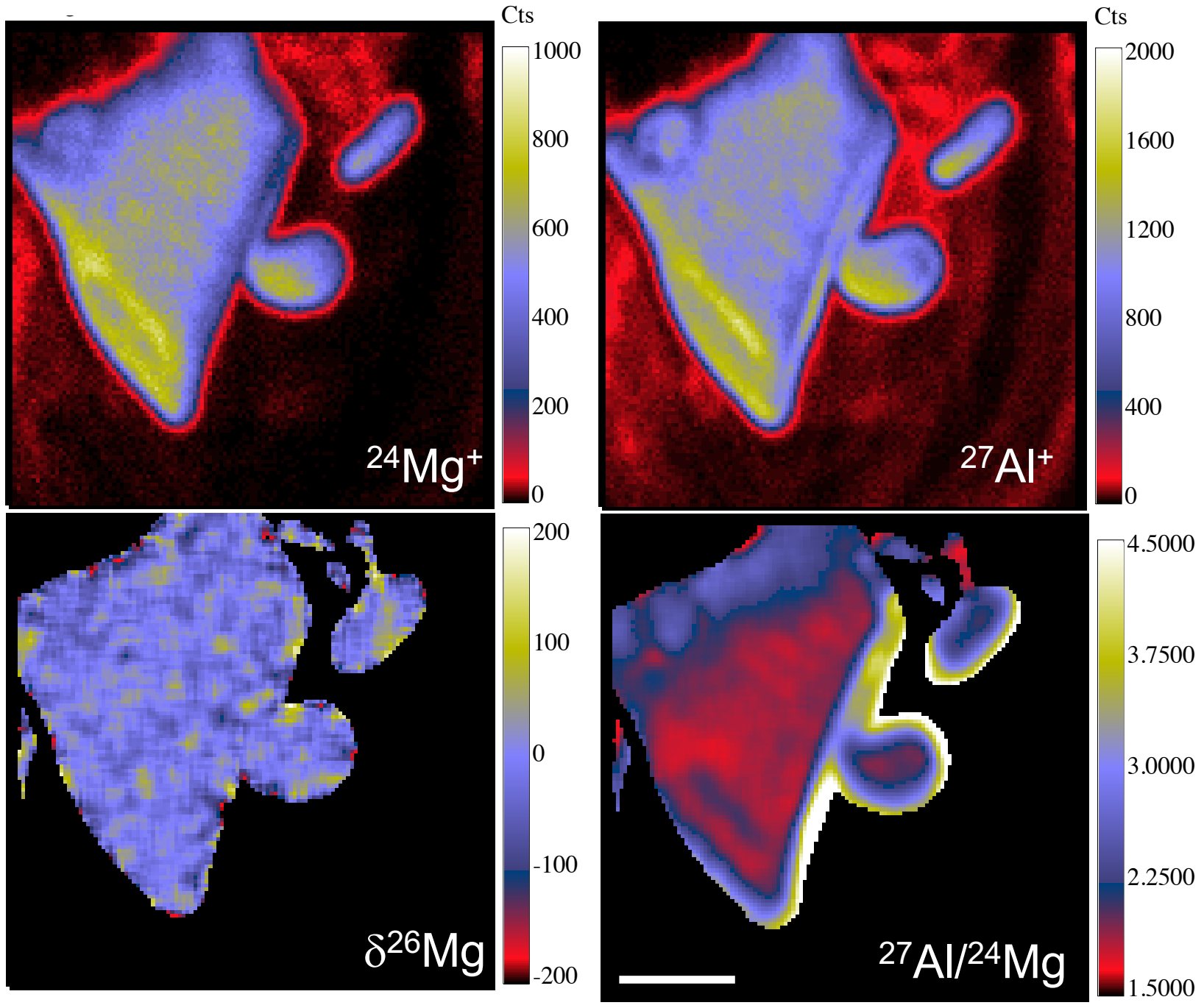

**Figure A5**. *NanoSIMS Images of Burma spinel grains. The white scalebar denotes 1 μm.*

In this study, we investigated possibility (ii) by analyzing Burma spinel grains of varying sizes (0.5−2 μm) for their Mg-Al isotopes with NanoSIMS, and an example of the results is shown in Fig. A5 for illustration. Figure A5 reveals that the Al/Mg ratio determined by NanoSIMS analyses is significantly affected by an edge effect, i.e., increased Al/Mg ratios at grain rims, which is absent in the $^{26}$Mg/$^{24}$Mg ratio image. Because of this edge effect, the measured $^{27}$Al/$^{24}$Mg ratio increases from 2.0 to 2.6 for the largest (2.2 μm) and smallest (0.6 μm) grains in Fig. A5, respectively, corresponding to a 30% decrease in $\Gamma_{Mg/Al}$ with decreasing grain size. It is, however, difficult to quantity the edge effect since it depends on the edge topography that varies on a case-by-case basis. A practical approach to suppress the edge effect is to exclude grain edges from the chosen ROIs for data reduction, which can be done for grains >500 nm given the thickness of the edge (>200 nm) and the spatial resolution of the NanoSIMS analyses (~100 nm) (Fig. A5). Excluding grain edges for data reduction is also necessary to suppress the effects of Al contamination rims on the derived initial $^{26}$Al/$^{27}$Al ratios for presolar SiC grains. Indeed, our derived $\Gamma_{Mg/Al}$ values based on polished NIST glass and large (>1 μm) Burma spinel grains are in good agreement within uncertainties (Section 2).

Based on the lack of dependence of the derived initial $^{26}Al/^{27}Al$ ratio on grain size among our X grains (Fig. A6), we conclude that the factor of two decrease in $\Gamma_{Mg/Al}$ in SiC is most likely a result of the reduced chemistry of SiC in comparison to NIST 610 and Burma spinel. Given the lack of proper standards for $Si_3N_4$ grains, we also adopted $\Gamma_{Mg/Al}$ = 0.69 for deriving the initial $^{26}Al/^{27}Al$ ratios of the four $Si_3N_4$ grains from this study.

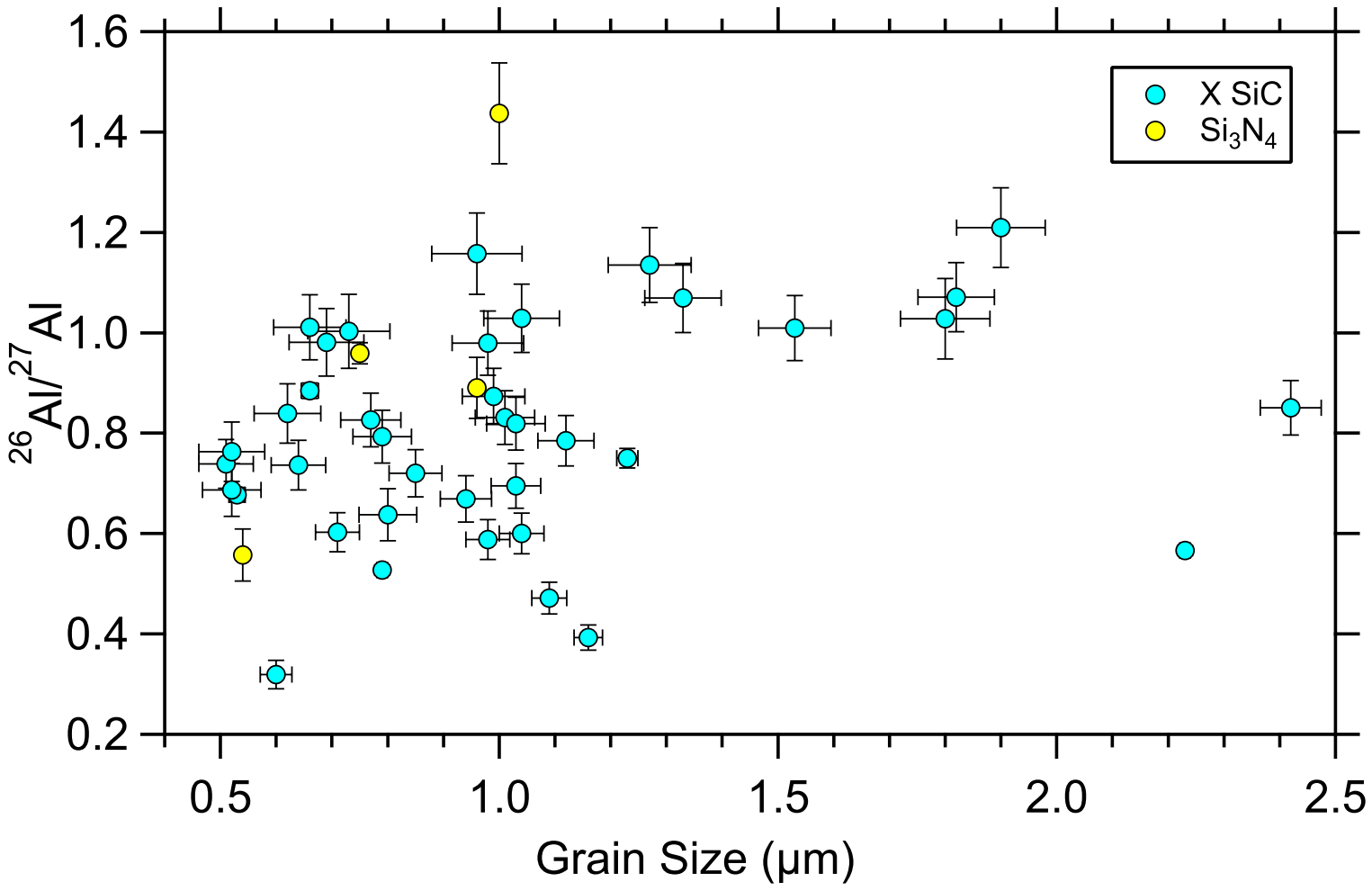

**Figure A6**. *Initial $^{26}Al/^{27}Al$ ratios of X and $Si_3N_4$ grains from this study as a function of grain size.*

## Appendix B

## Subtypes of Presolar X Grains

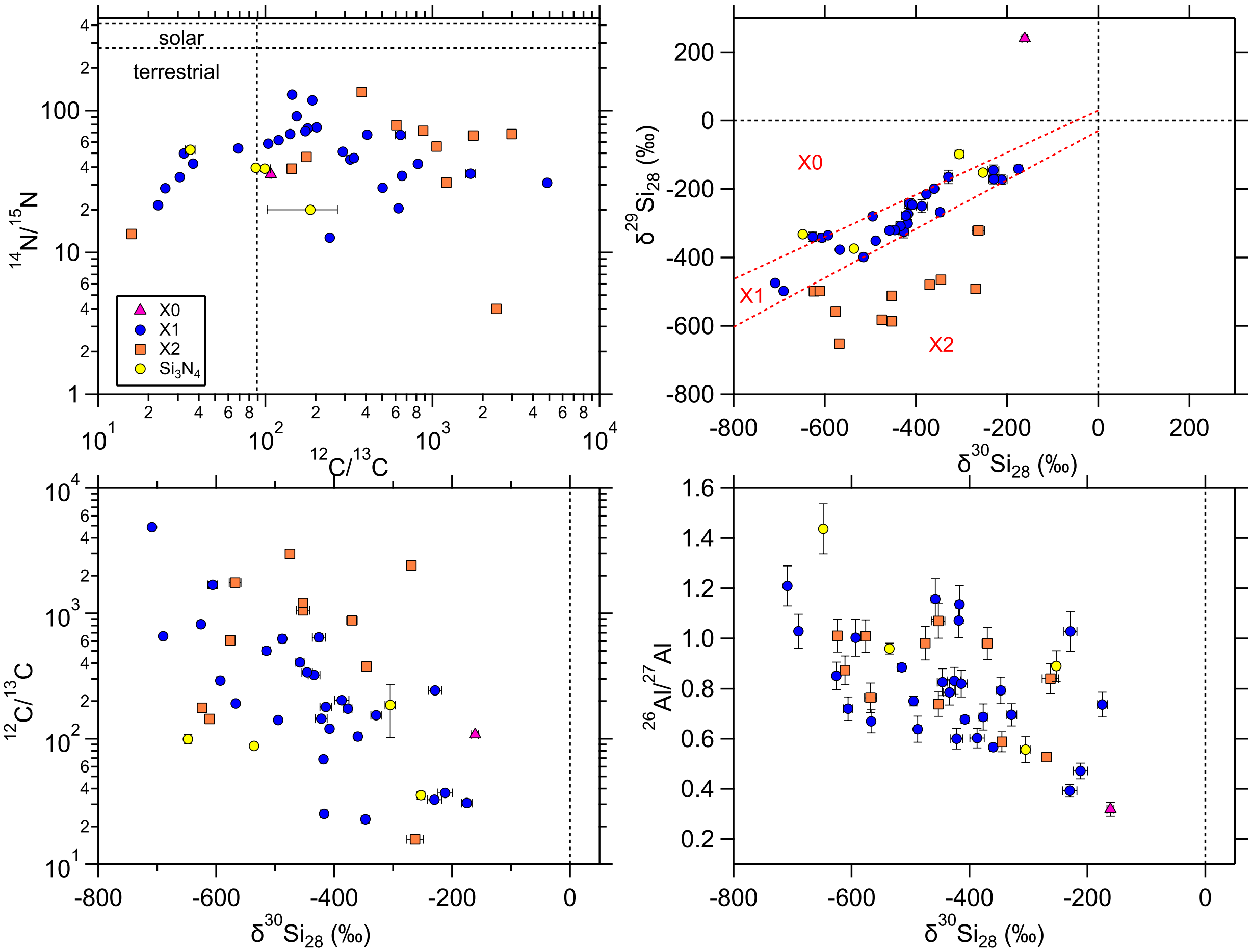

**Figure A7**. *Plots comparing the isotopic compositions of the three subtypes of X grains that were initially proposed by Lin et al. (2010) and later modified by Stephan et al. (2021).* $Si_3N_4$ *grain data are also plotted for comparison.*

# Appendix C

## Comparison of CCSN Grains with ν-models

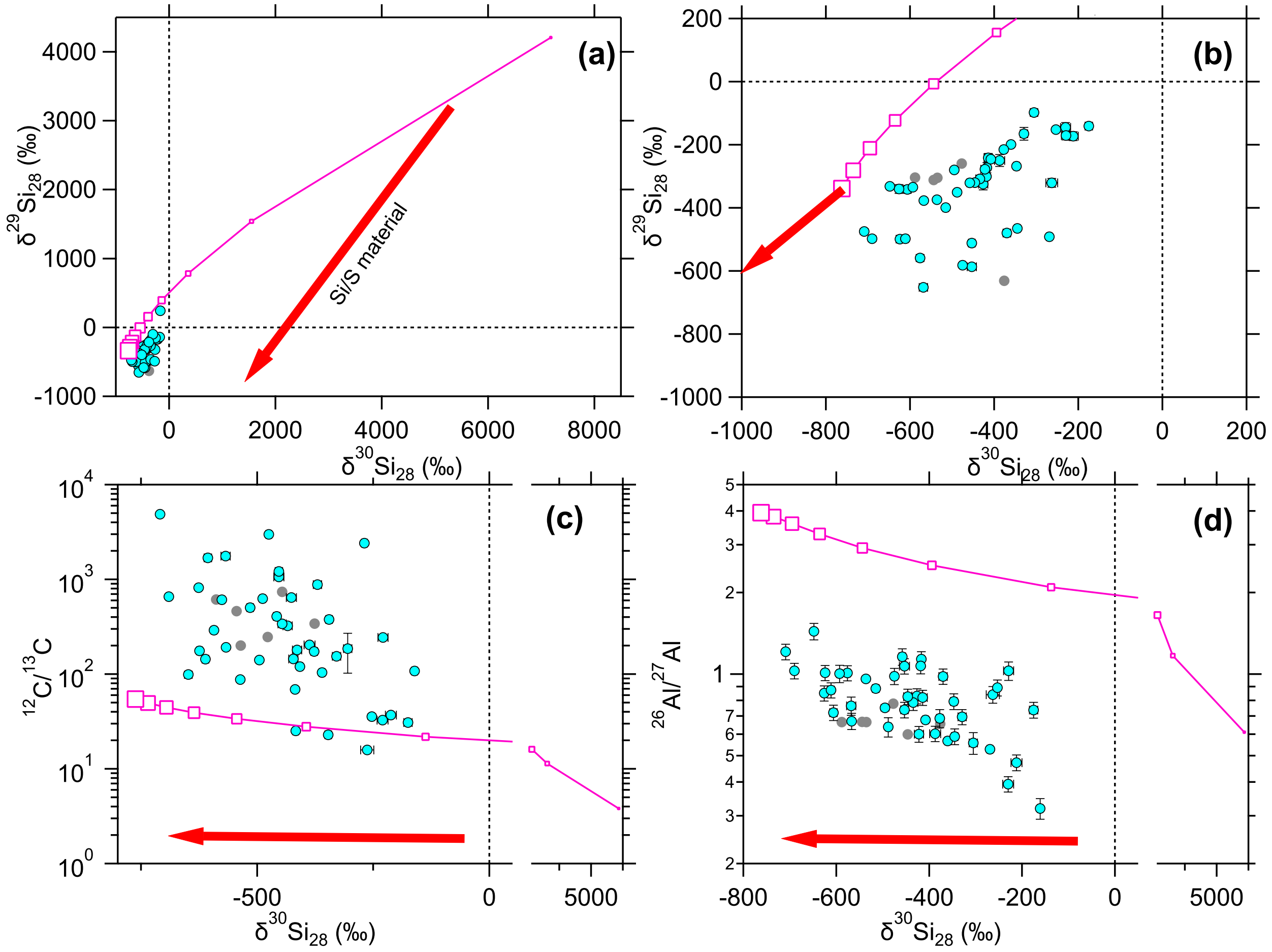

**Figure A8**. *Same as Fig. 3 except that the ν-model calculations for the Fe/Ni core are shown for comparisons at explosion energies of (1–10) ×10$^{51}$ erg.*

Table 1. Isotope and Elemental Ratio Data of 39 X SiC and four $Si_3N_4$ Grains from This Study (1σ errors).

| Grain | Phase | Subtype | Size (μm) | $^{12}C/^{13}C$ | $^{14}N/^{15}N$ | $\delta^{29}Si_{28}$ (‰)* | $\delta^{30}Si_{28}$ (‰) | $^{26}Al/^{27}Al$[#] | NanoSIMS[@] | | EDX | |
|---|---|---|---|---|---|---|---|---|---|---|---|---|
| | | | | | | | | | $^{26}Mg/Si$ (× $10^{-3}$) | $^{27}Al/Si$ (×$10^{-3}$) | $^{26}Mg/Si$ (× $10^{-3}$) | $^{27}Al/Si$ (× $10^{-3}$) |
| M5-A1-0522[&] | SiC | X1 | 2.2 | 103.9±0.9 | 58.6±1.3 | –199±5 | –360±6 | 0.57±0.01 | 6.0±0.6 | 18.7±1.4 | 3.0±0.7 | 8.1±0.9 |
| M5-A2-0022[&] | SiC | X1 | 0.7 | 203.2±8.6 | 76.4±1.2 | –250±19 | –387±12 | 0.60±0.04 | 8.4±0.9 | 29.9±3.9 | 6.6±1.6 | 9.6±1.6 |
| M5-A2-1614 | SiC | X2 | 0.6 | 15.8±0.2 | 13.5±0.2 | –321±14 | –263±14 | 0.84±0.06 | 5.6±0.6 | 14.2±1.9 | 4.7±1.8 | 7.2±2.0 |
| M5-A2-2885[&] | SiC | X2 | 1.5 | 608.6±27.6 | 79.0±1.3 | –559±11 | –576±6 | 1.01±0.06 | 11.4±1.2 | 24.1±3.2 | 11.8±1.0 | 16.4±1.0 |
| M5-A2-4297 | SiC | X1 | 1.2 | 32.6±0.4 | 49.9±1.1 | –145±15 | –230±12 | 0.39±0.03 | 7.2±0.8 | 39.4±5.2 | 3.9±0.8 | 9.7±1.0 |
| M5-A2-4351[&] | SiC | X2 | 0.7 | 176.3±6.0 | 47.2±0.8 | –499±12 | –624±6 | 1.01±0.06 | 26.6±2.9 | 56.5±7.4 | 19.1±1.3 | 27.6±1.4 |
| M5-A3-0082 | SiC | X1 | 1.0 | 154.2±5.0 | 91.4±3.0 | –165±20 | –329±9 | 0.70±0.04 | 6.44±0.7 | 19.8±2.69 | 4.9±0.8 | 7.2±0.8 |
| M5-A3-0132[&] | SiC | X1 | 1.0 | 643.8±44.9 | 67.4±1.3 | –325±18 | –426±11 | 0.83±0.05 | 26.6±2.9 | 68.5±0.9 | 16.4±1.1 | 23.5±1.2 |
| M5-A3-0420[&] | SiC | X1 | 1.3 | 25.2±0.6 | 28.3±0.5 | –273±16 | –417±6 | 1.14±0.07 | 2.9±0.3 | 5.4±0.7 | 2.8±0.8 | 6.4±0.8 |
| M5-A3-0946 | SiC | X2 | 1.3 | 1059.1±65.0 | 55.8±0.9 | –587±9 | –453±11 | 1.07±0.07 | 15.6±1.7 | 31.3±4.1 | 5.8±0.6 | 9.2±0.8 |
| M5-A4-0309[&] | SiC | X1 | 0.8 | 1690.3±108.4 | 35.9±0.5 | –342±12 | –606±8 | 0.72±0.05 | 15.4±1.7 | 45.8±6.0 | 6.6±0.8 | 11.9±1.0 |
| M5-A4-0322-2 | SiC | X1 | 1.8 | 68.8±1.2 | 54.2±0.7 | –301±7 | –418±5 | 1.07±0.07 | 9.9±1.1 | 19.8±2.6 | 6.3±0.6 | 9.2±0.8 |
| M5-A4-0361 | SiC | X1 | 1.0 | 179.4±4.0 | 75.0±1.0 | –242±14 | –414±10 | 0.82±0.05 | 17.5±1.9 | 45.6±6.0 | 30.6±1.2 | 44.9±1.5 |
| M5-A4-0691 | SiC | X1 | 1.1 | 322.3±8.5 | 45.2±0.6 | –308±12 | –434±10 | 0.79±0.05 | 10.0±1.1 | 27.8±3.7 | 9.5±1.0 | 13.4±1.0 |
| M5-A5-0071 | SiC | X1 | 2.4 | 818.3±25.2 | 42.1±0.6 | –340±14 | –626±5 | 0.85±0.05 | 7.9±0.9 | 19.8±2.6 | 11.7±4.6 | 9.1±3.7 |
| M5-A5-2856[&] | SiC | X2 | 1.0 | 143.7±2.7 | 39.0±0.5 | –498±9 | –611±7 | 0.87±0.06 | 39.8±4.3 | 97.6±12.8 | 30.2±13.2 | 31.3±13.9 |
| M5-A6-1154 | $Si_3N_4$ | X1 | 0.8 | 87.6±3.1 | 39.5±0.8 | –374±3 | –536±4 | 0.96±0.02 | 3.4±0.4 | 6.2±0.5 | 15.4±1.7 | 38.4±2.3 |
| M5-A7-0303 | SiC | X2 | 1.0 | 878.8±25.8 | 72.0±1.0 | –480±9 | –370±9 | 0.98±0.06 | 7.6±0.8 | 16.6±2.2 | 3.7±0.8 | 4.8±0.8 |
| M5-A7-0663[&] | SiC | X1 | 1.0 | 656.6±19.0 | 34.7±0.5 | –498±9 | –690±5 | 1.03±0.07 | 5.1±0.6 | 10.6±1.4 | 5.7±0.9 | 8.9±0.9 |
| M5-A7-1003 | SiC | X1 | 0.8 | 338.4±8.5 | 46.4±0.7 | –320±12 | –446±9 | 0.83±0.05 | 7.7±0.8 | 20.0±2.6 | 3.5±1.0 | 29.2±1.4 |
| M5-A7-1613[&] | SiC | X1 | 1.1 | 37.0±0.6 | 42.3±0.5 | –173±14 | –212±12 | 0.47±0.03 | 8.2±0.9 | 37.0±4.9 | 4.4±1.0 | 10.6±1.3 |
| M5-A7-2335[&] | SiC | X1 | 1.8 | 243.0±4.8 | 12.7±0.2 | –171±14 | –229±11 | 1.03±0.08 | 0.4±0.1 | 0.9±0.1 | N.D.[$] | N.D. |
| M5-A7-2884[&] | SiC | X1 | 1.9 | 4862.0±182.4 | 31.0±0.4 | –475±9 | –709±4 | 1.21±0.08 | 3.6±0.4 | 6.4±0.8 | 3.3±0.5 | 5.2±0.5 |
| M5-A7-2897 | SiC | X2 | 0.5 | 1212.3±36.2 | 31.1±0.2 | –512±8 | –453±8 | 0.74±0.05 | 3.8±0.4 | 11.1±1.5 | 3.7±0.8 | 6.5±1.0 |
| M6-A1-0228[&] | SiC | X1 | 0.6 | 30.8±0.2 | 33.9±0.8 | –141±7 | –175±9 | 0.74±0.05 | 28.8±3.1 | 77.9±6.0 | 22.1±1.9 | 178.4±2.9 |
| M6-A1-4635 | SiC | X1 | 0.8 | 22.8±0.1 | 21.5±0.5 | –268±6 | –347±7 | 0.79±0.05 | 12.4±1.4 | 31.3±2.4 | 22.1±1.7 | 117.0±2.2 |
| M6-A1-5959[&] | SiC | X1 | 0.5 | 120.1±1.2 | 62.0±1.4 | –246±5 | –408±6 | 0.68±0.01 | 10.9±1.2 | 28.3±2.2 | 53.9±1.8 | 6.4±2.2 |
| M6-A2-0222[&] | SiC | X1 | 1.2 | 140.9±1.1 | 68.3±1.6 | –280±3 | –495±4 | 0.75±0.02 | 2.9±0.3 | 6.8±0.5 | 1.8±0.9 | 3.9±0.9 |
| M6-A3-0852 | SiC | X1 | 0.7 | 144.4±2.2 | 129.7±4.7 | –278±10 | –422±10 | 0.60±0.04 | 8.6±0.9 | 28.7±2.2 | 5.8±1.2 | 10.5±1.2 |
| M6-A3-0983 | $Si_3N_4$ | X1 | 1.0 | 35.5±2.4 | 53.0±2.0 | –152±5 | –253±5 | 0.89±0.06 | 3.9±0.4 | 8.7±0.7 | N.D. | 9.5±1.4 |
| M6-A3-1169[&] | SiC | X1 | 0.7 | 291.1±6.6 | 51.3±1.7 | –335±6 | –593±6 | 1.00±0.07 | 5.4±0.6 | 10.6±0.9 | 8.0±1.4 | 12.9±1.7 |
| M6-A3-1802[&] | SiC | X2 | 0.5 | 1755.5±118.8 | 66.6±3.1 | –652±7 | –568±10 | 0.76±0.06 | 10.4±1.2 | 27.2±2.2 | 8.3±1.3 | 11.4±1.6 |
| M6-A4-0151[&] | SiC | X1 | 0.8 | 627.0±13.4 | 20.5±0.5 | –351±6 | –488±6 | 0.64±0.05 | 4.6±0.5 | 14.4±1.2 | 5.5±1.5 | 14.8±2.0 |
| M6-A5-0243 | SiC | X2 | 0.7 | 2974.7±151.7 | 68.4±2.6 | –582±5 | –475±6 | 0.98±0.07 | 4.1±0.4 | 8.3±0.6 | 4.7±1.0 | 22.0±0.5 |
| M6-A5-1485 | SiC | X1 | 1.0 | 406.5±12.3 | 67.6±2.5 | –321±6 | –458±7 | 1.16±0.08 | 11.6±1.3 | 19.9±1.6 | 12.8±1.1 | 70.5±1.1 |
| M6-A5-1858 | SiC | X1 | 0.9 | 191.2±2.4 | 118.2±4.3 | –377±4 | –567±4 | 0.67±0.05 | 2.8±0.3 | 8.4±0.6 | 4.1±0.9 | 31.1±0.5 |
| M6-A6-0109 | SiC | X2 | 1.0 | 377.7±7.9 | 135.2±4.8 | –465±6 | –345±7 | 0.59±0.04 | 12.2±1.3 | 41.3±3.2 | 10.7±1.6 | 64.1±0.9 |
| M6-A6-1330[&] | $Si_3N_4$ | X0 | 1.0 | 99.2±8.5 | 39.0±1.0 | –332±5 | –648±4 | 1.41±0.10 | 1.9±0.2 | 2.6±0.2 | 3.7±1.7 | N.D. |
| M6-A6-3269[&] | $Si_3N_4$ | X0 | 0.5 | 185.8±83.5 | 20.2±0.5 | –98±9 | –305±9 | 0.56±0.05 | 1.6±0.2 | 5.9±0.5 | 7.1±3.3 | 11.1±3.3 |
| M6-A7-0883 | SiC | X1 | 0.7 | 503.2±12.9 | 28.5±0.5 | –399±6 | –515±7 | 0.88±0.02 | 87.4±9.5 | 173.9±13.5 | 31.5±1.7 | 39.1±1.8 |
| M6-A7-2790-2 | SiC | X1 | 0.5 | 173.4±2.5 | 71.6±1.2 | –215±7 | –377±7 | 0.69±0.05 | 24.8±2.7 | 72.4±5.6 | 2.7±0.8 | 2.2±0.8 |
| M6-A7-3156[&] | SiC | X0 | 0.6 | 107.7±1.2 | 35.5±0.6 | 240±7 | –161±6 | 0.32±0.03 | 1.3±0.1 | 8.0±0.6 | 2.6±1.3 | 7.6±1.4 |
| M6-A7-3184[&] | SiC | X2 | 0.8 | 2411.7±88.3 | 4.0±0.1 | –492±4 | –269±6 | 0.53±0.01 | 20.3±2.2 | 67.9±5.3 | 12.2±1.3 | 18.9±1.5 |

[*]: $\delta^{i}Si_{28}$ was calculated using the equation $\delta^{i}Si_{28} = [(^{i}Si/^{28}Si)_{grain} / (^{i}Si/^{28}Si)_{std} - 1] \times 1000$, in which $^{i}Si$ denotes $^{29}Si$ or $^{30}Si$.

[#]: The initial $^{26}Al/^{27}Al$ ratios were calculated by adopting the $\Gamma_{Mg/Al}$ value (0.69) inferred in Section 3.1. The associated errors are based on counting statistics and uncertainties in $\Gamma_{Mg/Al}$ determined from Burma spinel measurements.

[@]: The associated errors are based on counting statistics and uncertainties in $\Gamma_{Mg/Si}$ and $\Gamma_{Al/Si}$ values determined from NIST 610 measurements.

[&]: This X grain had relatively uniform $^{26}Mg$ and was thus included in Fig. 2a to derive $\Gamma_{Mg/Si}$ value.

[$]: N.D. stands for not detected.